\documentstyle[12pt]{article}                                                   
\newcommand{\njl}{Nambu-Jona-Lasinio}                                          
\def\del{\partial}

\begin{document}                                                                
                                                                                
\begin{titlepage}                                                               
December, 1993 \hfill                 BNL-ABG-1
\begin{center}                                                                  
{\LARGE\bf Spontaneous Breaking of Flavor Symmetry and Parity in the 
Nambu-Jona-Lasinio Model with Wilson Fermions}

\vspace{1.5cm}                                                                  
\large{
S. Aoki\\ 
Institute of Physics, University of Tsukuba\\
Tsukuba, Ibaraki 305, Japan\\
}
\vspace{0.5cm}
\large{                                                                  
and\\
}
\vspace{0.5cm}
\large{
S. Boettcher and A. Gocksch\footnote{Address after December 27 1993:
Morgan Stanley \& CO. Incorporated, 1221 Avenue of the Americas, New York, NY
10020.}\\
Physics Department, Brookhaven National Laboratory\\
P.O. Box 5000\\
Upton, NY 11973, USA \\
}
\vspace{1.9cm}                                                                  
\end{center}                                                                    

\abstract{
We study the lattice \njl~model with two flavors of Wilson fermions in the
large $N$ limit, where $N$ is the number of `colors'. For large values
of the four-fermion coupling we find a phase in which both, flavor symmetry
and parity, are spontaneously broken. In accordance with general expectations
there are three massless pions on the phase boundary, but only two of them
remain massless inside the broken phase. This is analogous to earlier results
obtained in lattice QCD, indicating that this behavior is a very general 
feature of the Wilson term.
}
\vfill                                                                          
\end{titlepage}
\flushbottom
\section*{1. INTRODUCTION}
Recently, Bitar and Vranas\cite{bitar} presented an extensive study of
the lattice Nambu-Jona-Lasinio model. This model, although interesting
in its own right as an effective low-energy theory of the strong interactions,
can serve as an ideal testing ground for the properties of the Wilson term.
This is because the model can be easily studied in the large $N$ limit,
where in this case $N$ refers to the number of colors. In this letter we
are concerned primarily with the nature of the symmetry breaking induced by the
Wilson term. We will show that in the large $N$ limit there exists a phase
in which both parity and flavor symmetry are spontaneously broken. It is
the existence of this phase transition that is responsible for the occurrence
of massless pions in the model despite the fact that chiral symmetry is
{\it explicitly} broken by the Wilson term. The same phenomenon is also
responsible for the masslessness of the pion in lattice QCD\cite{sinya}.
The existence of this phase was
missed by Bitar and Vranas. Nevertheless, most of our results in the symmetric
phase are in agreement with theirs.

The model under consideration is defined through the action
\begin{equation}
S=\sum_{x,y}\sum_{i=1}^N \Biggl\{\bar \psi^i (x)[ M(x,y)+M^{\dagger}(x,y)] 
\psi^i (y) +{2\beta \over N}\bigl[\sigma^2(x)+\pi^2(x)\bigr]\delta(x,y)\Biggr\},
\end{equation}
where $\psi^i(x)$ has also two flavor components in addition to the indices
shown. The combination $\beta \over N$ will be refered to as
$\tilde \beta$ below. The matrix $M(x,y)$ is defined by
\begin{eqnarray}
M(x,y)={1 \over 2}\sum_{\mu}\biggl[ (\gamma_{\mu}-r)\delta(x+\mu,y)
-(\gamma_{\mu}+r)\delta(x-\mu,y)\biggr]& \nonumber \\
+\delta(x,y)\biggl[4r+m+\sigma(x)+i\gamma_5 \vec \pi(x)\cdot \vec \tau\biggr]&.
\end{eqnarray}
In the above action auxiliary fields $\pi$ and $\sigma$ have been introduced
to decouple the four fermion interaction\cite{bitar}. The terms proportional
to the parameter $r$ come from the Wilson term. In the following we will 
set $r=1$. The Wilson term is of order $O(a)$ ($a$ is the lattice spacing) 
in the naive continuum limit but nevertheless it has a pronounced effect on 
the theory. It gives the fermionic doublers masses on the order of the 
momentum space cutoff, $O({r \over a})$, and in QCD it also serves to produce 
the correct anomaly of the flavor singlet axial current in the continuum 
limit\cite{smit}.  Note that the Wilson term explicitly breaks chiral symmetry.

As it is well known, the \njl~ model in the absence of the Wilson term breaks
chiral symmetry spontaneously at a critical value $\beta_c$. Above $\beta_c$,
the theory is chirally symmetric for $m=0$, and $<\bar \psi^i  \psi^i> $
serves as an order parameter for the transition.
With the Wilson term in place however, $<\bar \psi^i  \psi^i> $ is always
non-zero and chiral symmetry is explicitly broken. Massless pions,
if they exist in the model, therefore cannot considered to be 
the Goldstone bosons of broken chiral symmetry. We will show below
that massless pions do exist and that they must considered to be
Goldstone bosons of spontaneously broken {\it flavor} symmetry. In
addition to the Goldstone pions there is one remaining pion which
is massless {\it only} on the phase boundary and is associated with
the spontaneous breaking of a discrete space-time symmetry, parity.
In general, when there are $n_f$ flavors, there are ${n_f}^2-2$
Goldstone bosons and one mode which is massless only on the phase 
boundary. The flavor singlet pseudo-scalar meson is always massive.

In Sec.~2 we derive the formulas for the condensate and the
masses in the large $N$ limit. In Sec.~3 we present the results of our
calculation, and finally Sec.~4 contains a short discussion of our results. 

\section*{2. LARGE N APPROXIMATION}
We start by integrating out the fermion fields in Eq. (1) obtaining
the following effective action for the auxiliary fields:
\begin{equation}
S=-{N \over 2}\biggl[Tr(\log M)+Tr(\log M^\dagger)\biggr]
 +{2N\tilde \beta}\sum_x \biggl[\sigma^2(x)+\pi^2(x)\biggr],
\end{equation}
where the trace extends over space, color, flavor and spin degrees of
freedom. Since the action is proportional to $N$ we can proceed in 
the large $N$ limit by evaluating the functional integral around the
stationary point of the action. We write
\begin{equation}
\sigma(x) \sim \sigma_s+{\delta \sigma (x) \over \sqrt N},\qquad
\vec \pi(x) \sim \pi_s \hat e_3+{\delta \vec \pi (x) \over \sqrt N},\qquad
N\to\infty,
\end{equation} 
where we have accounted for fluctuations around the translationally
invariant saddle point. Note, that we have allowed for the possibility
that the isovector $\vec \pi$ might develop a non-zero condensate
which we have arbitrarily chosen to point in the $3$-direction in
isospin space. Such a condensate of course breaks both, parity and
flavor symmetry.

In momentum space, the inverse propagator at the saddle point is given by
\begin{equation}
\tilde M_s (p)=i \sum_{\mu} \gamma_{\mu} \sin p_{\mu}+(4-\sum_{\mu}\cos p_{\mu}
)+m+\sigma_s+i \gamma_5 \tau_3 \pi_s.
\end{equation} 
The effective potential evaluated at the saddle point is
\begin{equation}
V_{eff}=-2N n_f\int{d^4p \over (2 \pi)^4}\log[g(p)]+N n_f \tilde \beta 
({\sigma^2}_s+{\pi^2}_s)
\end{equation} 
where the function $g(p)$ is defined as
\begin{equation}
g(p)=\sum_{\mu}\sin^2 p_{\mu}+\pi_s^2 +[w(p)+m_q]^2.
\end{equation} 
Following Bitar and Vranas\cite{bitar} we define the constituent quark
mass $m_q=m+\sigma_s$ and the function $w(p)=4-\sum_{\mu}\cos p_{\mu}$. 
Demanding that the linear term in the fluctuations around the saddle 
point vanish, leads to the two gap equations
\begin{eqnarray}
0&=&\sigma_s{\tilde \beta\over 2}-
\int{d^4p \over (2 \pi)^4}{\sigma_s + m +w(p) \over g(p)}, \\
0&=&\pi_s{\tilde \beta\over 2}-\int{d^4p \over (2 \pi)^4}{\pi_s \over g(p)}.
\end{eqnarray}  
Obviously, $\pi_s=0$ is always a solution to the last equation, but as it 
turns out it is not always the one that minimizes the effective potential. 

The inverse propagators are obtained from the quadratic fluctuations
and are given by
\begin{eqnarray}
&&G_{\sigma}^{-1}(k)=\\
&&{\tilde \beta \over 2} -\int{d^4p \over (2 \pi)^4}
{\sum_{\mu}\sin (p_{\mu}+{k_{\mu}\over 2})\sin (p_{\mu}-{k_{\mu}\over 2})
+\pi_s^2 -[w(p+{k\over 2})+m_q][w(p-{k\over 2})+m_q] 
\over g(p+{k\over2})g(p-{k\over2})},\nonumber 
\end{eqnarray} 
and
\begin{eqnarray}
&&G_{\pi^a}^{-1}(k)=2\pi_s^2\delta^{a,3}\int{d^4p\over (2\pi)^4}
{1\over g(p+{k\over2})g(p-{k\over2})} \\
&&+{\tilde \beta \over 2} - \int{d^4p \over (2 \pi)^4}
{\sum_{\mu}\sin (p_{\mu}+{k_{\mu}\over 2})\sin (p_{\mu}-{k_{\mu}\over 2})
+\pi_s^2 +[w(p+{k\over 2})+m_q][w(p-{k\over 2})+m_q] 
\over g(p+{k\over2})g(p-{k\over2})}. \nonumber 
\end{eqnarray} 
In the last equation we have used that the pion propagator is diagonal
in flavor space. Note, that $G_{\pi^a}^{-1}(0)=0$ for $a=1,~2$ due to 
Eq. (9) when $\pi_s\not=0$. 

It is very difficult to calculate the mass, $m_\sigma$, for the
pseudo-scalar meson from the complex zeros of the inverse
propagator in Eq. (10)\cite{bitar}. To obtain the mass of the pions, which is
expected to be small close to the phase boundary, we can define the pion
wave function renormalization constant $Z_{\pi^a}$ and pion mass $m_{\pi^a}$
such that
\begin{equation}
\lim_{k\to 0} G_{\pi^a}^{-1}(k) = Z_{\pi^a}^{-1} \left( k^2+m_{\pi^a}^2\right).
\end{equation}
Unless $Z_{\pi^a}$ becomes infinite, the pion masses can be computed from
\begin{equation}
m_{\pi^a}^2=\lim_{k\to 0} { G_{\pi^a}^{-1}(k) \over 
{\del\over \del k^2} G_{\pi^a}^{-1}(k) }.
\end{equation}
Note that since parity is broken in the phase in which $\pi_s \not= 0$,
the mixed $\pi-\sigma$ propagator is nonvanishing. Physical states in this
phase are not eigenstates of parity and are obtained by diagonalizing the 
mass matrix.

\section*{3. RESULTS}
To make our results comparable with those presented in \cite{bitar}, we
choose $N=2$ so that $\beta=2\tilde\beta$ in the following.
We solved the gap equations (8-9) numerically on a $10^4$-lattice 
using a simple Newton procedure with the bare parameters ranging form 
$0<\beta\leq2.5$ and $-2\leq m\leq -6$, each in steps of $0.01$. We compared 
the results of the calculation where $\pi_s$ is allowed to take a 
non-zero solution to those results that are obtained for $\pi_s$ 
set to zero. We found throughout that $\pi\not=0$ usually minimizes the 
effective potential in Eq. (6), if such a solution exists. Some care must 
be taken in this calculation due to threefold solutions of the gap 
equation for $\sigma_s$ when $\pi_s=0$ and $\beta<0.75$. In all cases we 
determined the solutions for $\pi_s$ and $\sigma_s$ that minimize the 
effective potential. The results for $\sigma_s$ and $\pi_s$ from this 
calculations are plotted in Figs.~1 and~2.

In Fig.~3, the continuous line represents the phase boundary between the 
regions where $\pi_s=0$ and the region where parity-flavor symmetry is 
broken and $\pi_s\not=0$. Note that the region with $\pi_s\not=0$ 
disappears for $\beta>1.41$. We have extended the calculation for $m=-4$ 
far into the weak-coupling regime where $\beta$ is large but there do 
not appear to be any more non-zero solutions for $\pi_s$. Along the 
dashed line in Fig.~3, the mass $m_q$ of the constituent quark vanishes.
The intersection of the line where $m_q=0$ with the phase boundary at 
$\beta\approx 0.33$ and $m\approx -2.7$ is the continuum chiral limit 
and corresponds reasonably well with the prediction of \cite{bitar}.

Close to the phase boundary the pion masses are expected to be small. Thus,
they can be calculated using Eq. (13). We find that the masses for all
three pions are equal on the side of the phase boundary where $\pi_s=0$
[as can be seen directly from Eq. (11)].
As one approaches the phase boundary we find that ${m_{\pi}}^2 \sim m-m_c$.
Although this is of course the behavior expected of a theory which breaks
chiral symmetry spontaneously, the reader should understand that here it 
is simply a consequence of the fact that the critical exponent of
the flavor-parity breaking transition has its mean field value at large $N$. 
In the broken phase, the masses of the $\pi^{1,2}$ remain zero, they are the
Goldstone modes corresponding to the two unbroken generators of flavor
symmetry. The mass of $\pi^3$ on the other hand is zero only at the
critical point. It is to be considered as the inverse correlation length
of an Ising like parity breaking transition. This behavior of the pion 
masses is summarized in Fig.~4. As we mentioned before, we were not able to 
obtain any useful results for the sigma mass. Its expected behavior
is such that it is nonvanishing throughout the phase diagram\cite{sinya}.

Finally, we would like to mention that we were unable to find
massless pions for values of $\beta > 1.41$. This seems to contradict
what was found in~ \cite{bitar} (see in particular Fig.~8 of this 
reference). On the basis of our interpretation of the phase
structure of the model one does not expect a massless particle
in that region since there are no phase transitions there.

\section*{4. CONCLUSION}
We have analyzed the phase structure of the \njl~model on the lattice
with Wilson fermions in the large $N$ limit. We found that in analogy
to lattice QCD with Wilson fermions there exists a phase in which flavor 
and parity symmetry are spontaneously broken. As opposed to what is
expected to happen in lattice QCD, the region of broken symmetry is
a simply connected, compact region, symmetric around $m=-4$. There appears
to be {\it no} interesting phase structure at weak coupling. 
Since a complicated phase structure has been found at weak coupling
of the lattice Gross-Neveu model (an asymptotically free
 four-fermi model in 2 dimensions) with Wilson fermions\cite{GN}, 
the absence of such an interesting phase structure at weak coupling
may be related to the fact that the Nambu-Jona-Lasino model
is non-asymptotic free.
We feel that
the most interesting aspect of our results is that one is able
to obtain both parity and flavor violation in a vector-like theory
in a well defined approximation scheme. Due to the presence of the
Wilson term, well known theorems\cite{vafa} about the absence of such
breaking do not apply. We hope that the results presented here will help
to make it easier to accept the parity-flavor breaking scenario for
lattice QCD as well. Also, in some of the recent proposals for putting
chiral fermions on the lattice, one is forced to work in a region where
$m<m_c$ (the broken region in our language)\cite{creutz}. A good understanding
of the nature of this phase is therefore very important.

\section*{ACKNOWLEDGEMENTS}
This manuscript was authored under Contract No. DE-AC02-76-CH00016
with the U.S. Department of Energy. Accordingly, the U.S. Government
retains a non-exclusive, royalty-free license to publish or reproduce
the published form of this contribution, or allow others to do so, for
U.S. Government purposes.

\newpage
\section*{FIGURE CAPTIONS}
\noindent
FIGURE 1: $\pi_s$ as function of the bare parameters $m$ and $\beta$,
obtained from the gap equations in (8-9). The value of $\pi_s$ is
non-zero only in a near-circular region (see Fig.~3) between
$0<\beta<1.41$ and $-5.5<m<-2.5$ that is symmetric with respect to the 
axis where $m=-4$. Note that for small $\beta$, $\pi_s$ appears to develop a 
singularity similar to $\exp\{-(4+m)^2/\beta\}/\sqrt\beta$.
\bigskip

\noindent
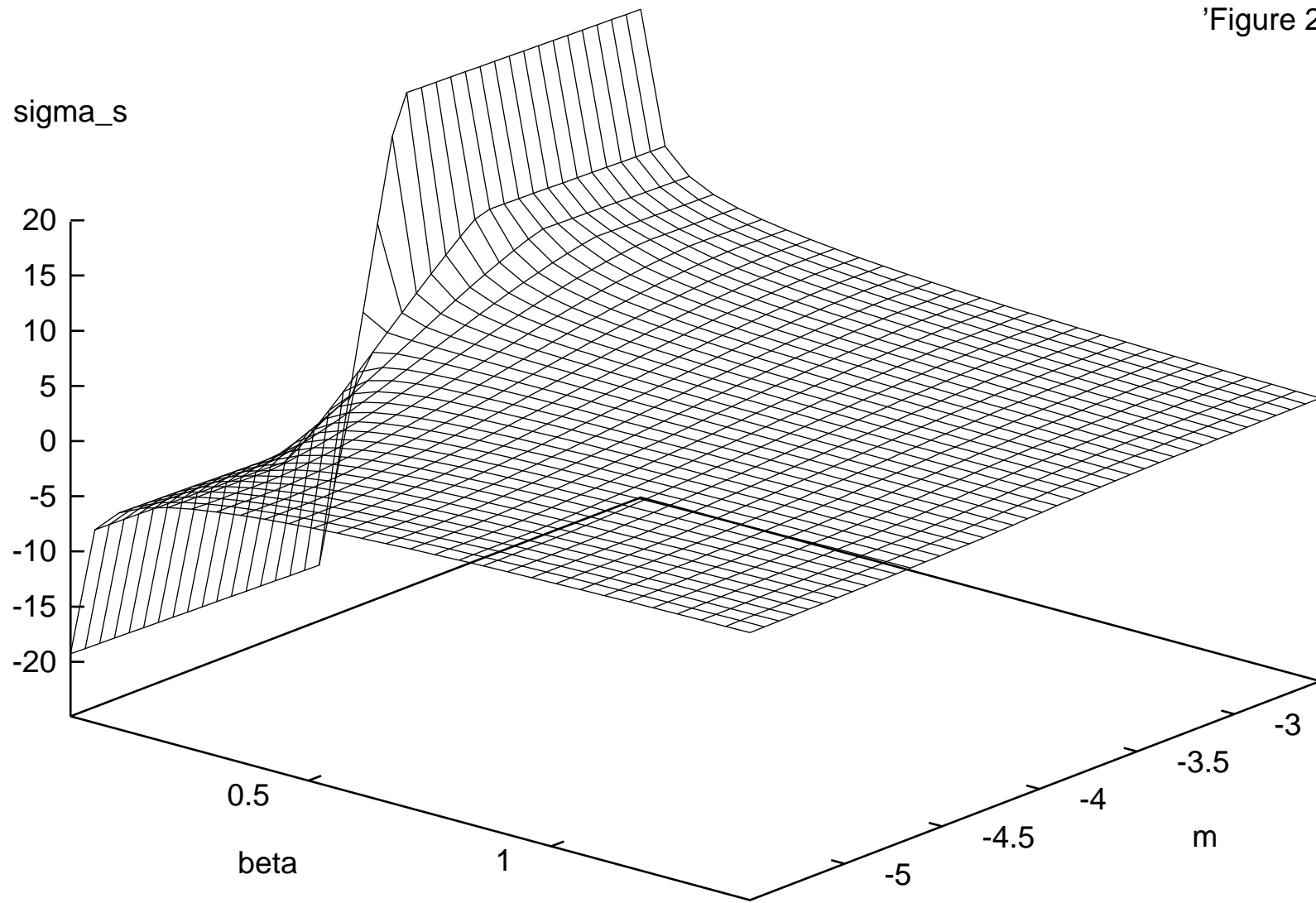
FIGURE 2: $\sigma_s$ as function of the bare parameters $m$ and $\beta$,
obtained from the gap equations in (8-9). The value of $\sigma_s$ is
antisymmetric with respect to the axis where $m=-4$, and passes continuously
through zero on this axis for all values of $\beta>0$.
\bigskip

\noindent
FIGURE 3:  Phase diagram for the \njl~model in Eq.~(1) with spontaneously 
broken parity-flavor symmetry. $\pi_s$ is non-zero only in the near-circular 
region that is symmetric with respect to the dotted line where $m=-4$. 
Everywhere outside of that region $\pi_s=0$, and the theory maintains the
symmetry. Along the dashed line the quark mass $m_q$ 
vanishes. The intersection of the line where $m_q=0$ with the phase
boundary where $m_\pi=0$ is the continuum chiral limit of the lattice theory.
\bigskip

\noindent
FIGURE 4: Pion masses near to the phase boundary for some generic
$\beta=0.58$. For $m>m_c=-2.804$, all pion masses are equal and non-zero.
The pion masses vanish for $m=m_c$, and, while ${m_{\pi^3}}^2$ becomes finite,
the masses for $\pi^{(1,2)}$ remain zero in the phase where parity-flavor 
symmetry is broken and $m<m_c$.

\newpage
\begin{thebibliography} {99}
\bibitem{bitar}
K. Bitar and P. Vranas, preprint FSU-SCRI-93-130 (October 1993) and
hep-lat/9311022.
\bibitem{smit}
L.H. Karsten and J. Smit, Nucl. \ Phys. \ B183 (1981) 103.
\bibitem{sinya}
S. Aoki, Phys. \ Rev. \ D30 (1984) 2653; 
Phys. \ Rev. \ D33 (1986) 2399; 
Phys. \ Rev. \ D34 (1986) 3170; 
Phys. \ Rev. \ Lett. \ 57 (1986) 3136; 
Phys. \ Lett. \ 190B (1987) 140; 
Nucl. \ Phys. \ B314 (1988) 79;
S. Aoki and A. Gocksch, Phys. \ Lett. \ 231B (1989) 449;
Phys. \ Lett. \ 243B (1990) 409; 
Phys. \ Rev. \ D45 (1992) 3845.
\bibitem{GN}
T. Eguchi and R. Nakayama, Phys. \ Lett. \ 185B (1981) 20;
S. Aoki and K. Higashijima, Prog. \ Theor. \ Phys. \ 76 (1986) 521.
\bibitem{vafa}
C. Vafa and E. Witten, Phys. \ Rev. \ Lett. \ 53 (1984) 535;
Nucl. \ Phys. \ B2 34(1984) 173.
\bibitem{creutz}
M. Creutz and I. Horvath, Poster at LATTICE '93, Dallas, TX, Oct. 12-16,
1993.
\end{thebibliography}
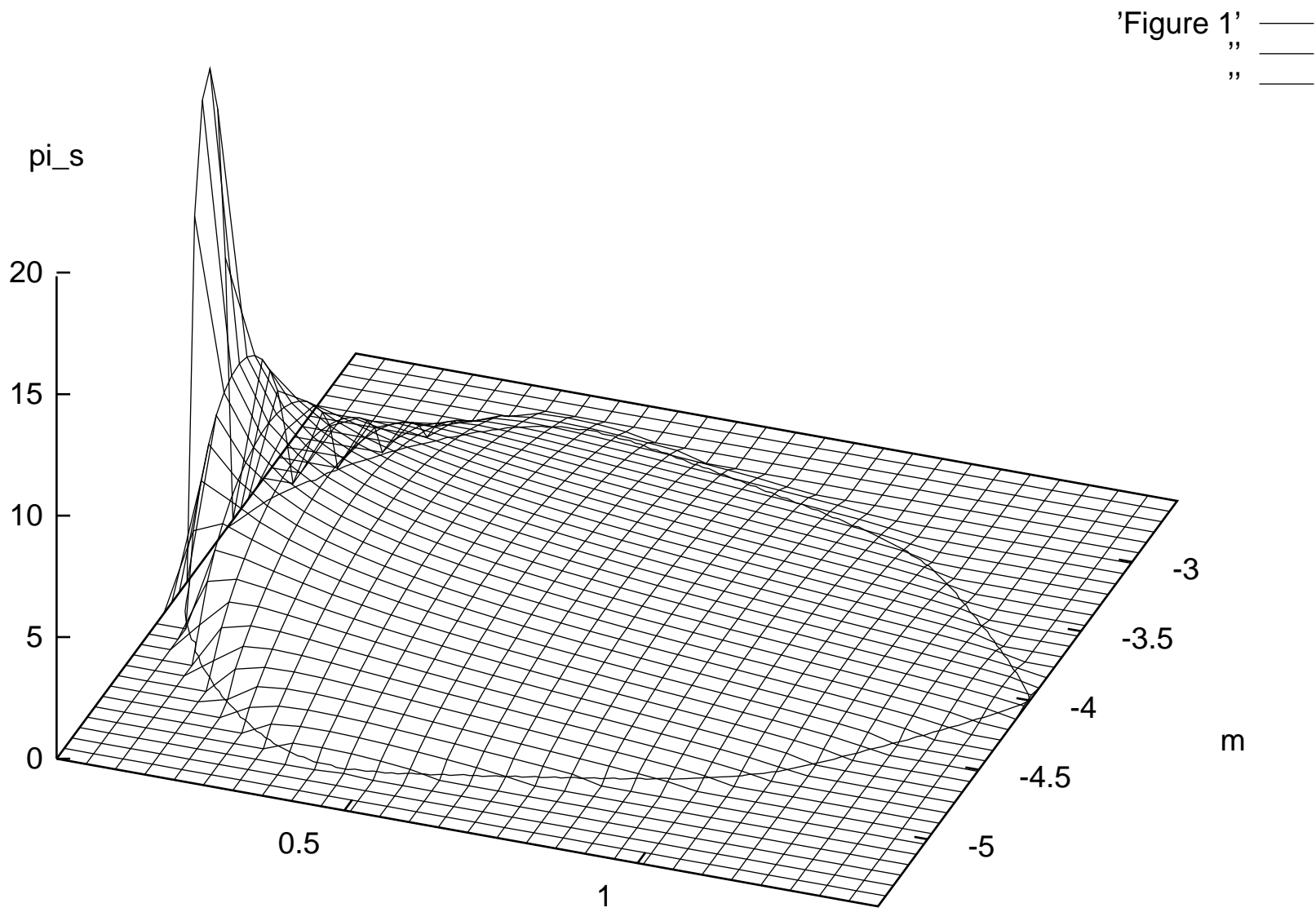
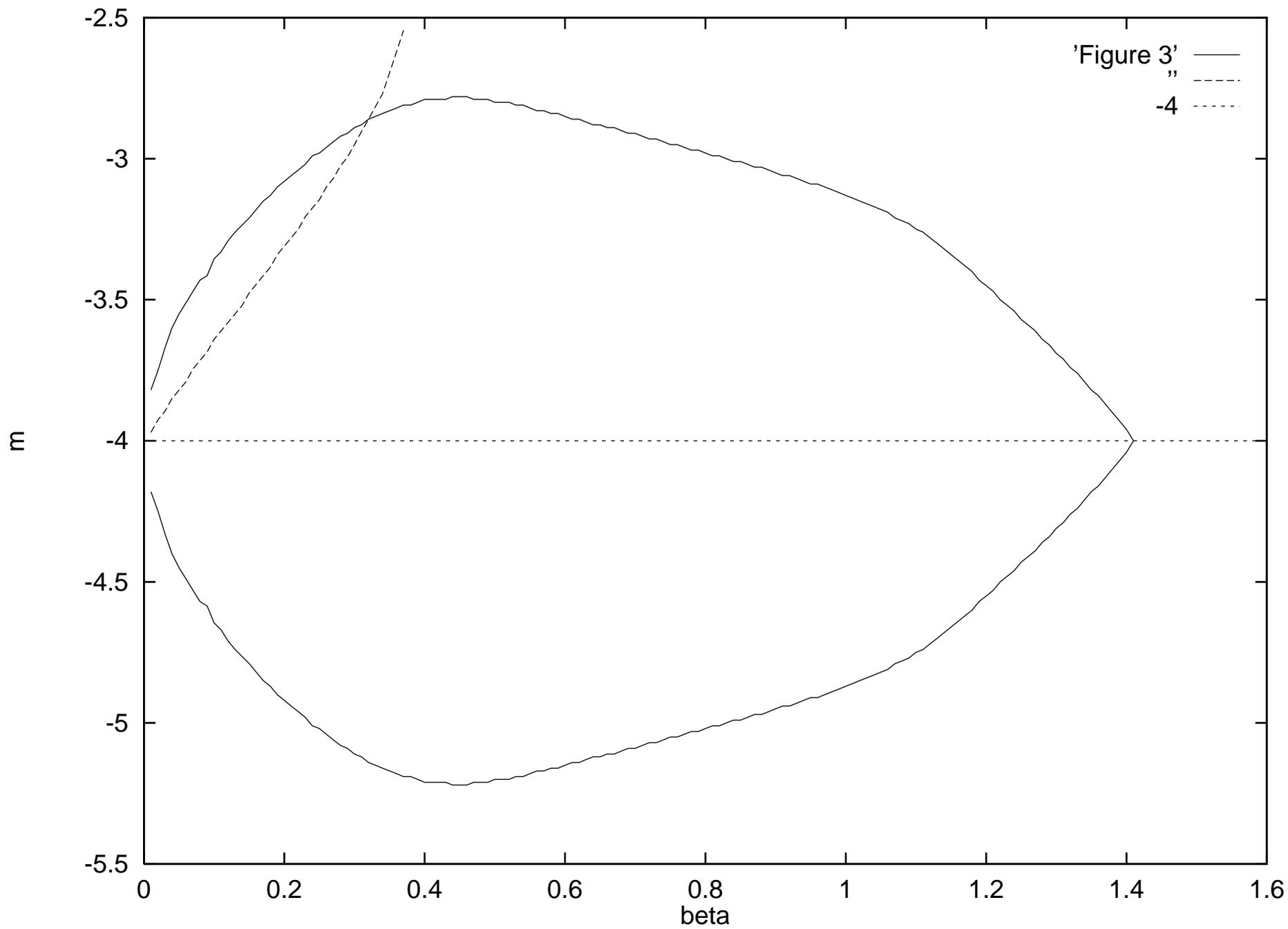
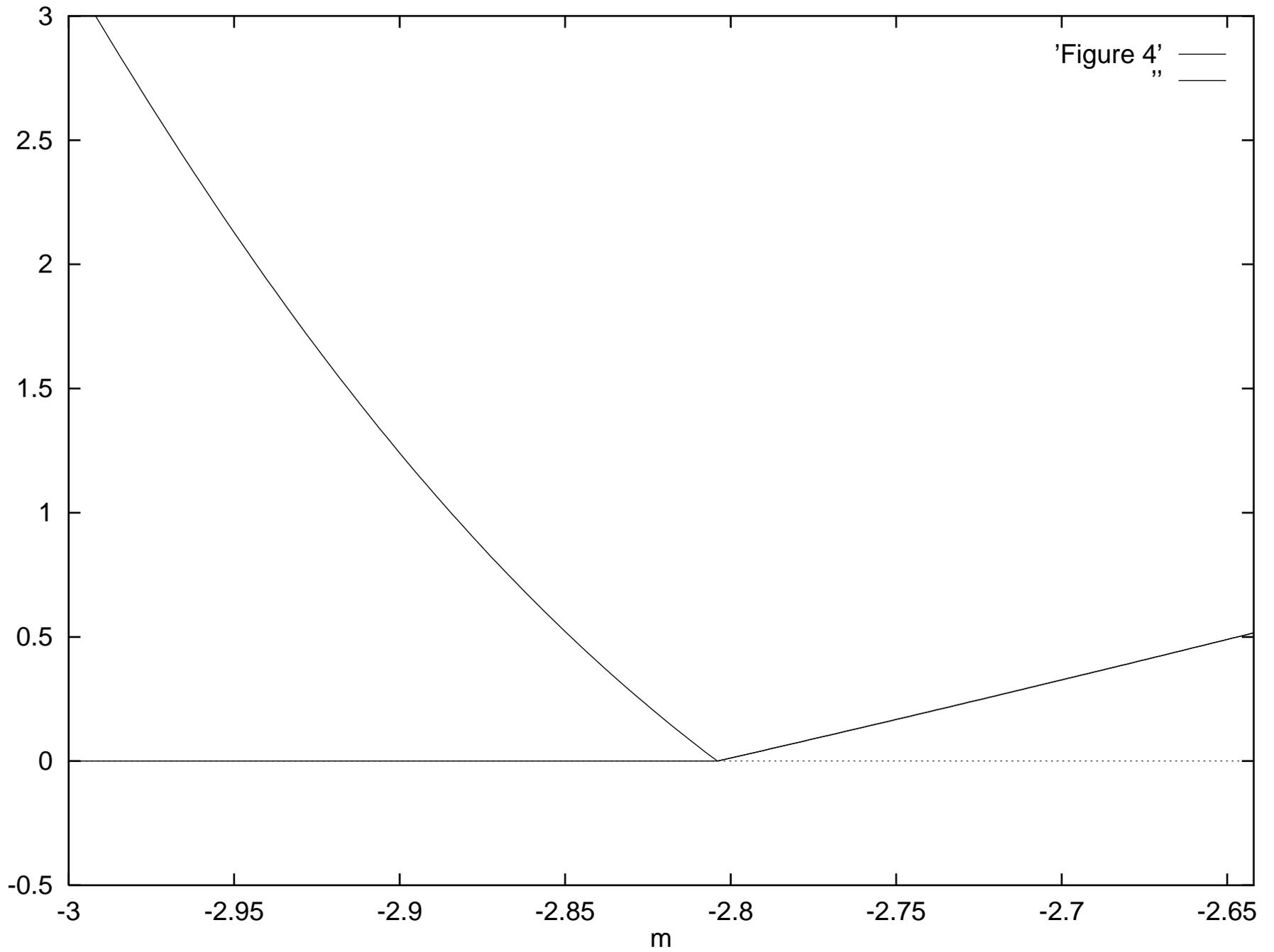
\end{document}